\documentclass[preprint,showpacs,preprintnumbers,amsmath,amssymb]{revtex4}
\usepackage{graphicx}
\usepackage{dcolumn}% Align table columns on decimal point
\usepackage{bm}% bold math

\begin{document}
%\draft
%\preprint{\today}

\title{ Re-appearance of the pairing correlations at finite temperature } 

%% Include following two lines for Journal Style

%\twocolumn[\hsize\textwidth\columnwidth\hsize  %**Journal
%\csname @twocolumnfalse\endcsname              %**Journal

\author{J.A. Sheikh$^{1}$, R. Palit$^{2}$ and S. Frauendorf$^{3}$}
\address{$^1$Department of Physics, University of Kashmir, Srinagar,
190 006, India \\
$^2$Tata Institute of Fundamental Research, Colaba, Mumbai, 400 005, India \\
$^{3}$Department of Physics, University of Notre Dame,
Notre Dame, IN 46556, USA
}

\date{\today}
%\maketitle

\begin{abstract}
Rotational and deformation dependence of isovector and isoscalar 
pairing correlations at finite temperature are studied in an exactly 
solvable cranked deformed shell model Hamiltonian. It is shown that isovector
pairing correlations, as expected, decrease with increasing deformation
and the isoscalar pairing correlations remain constant at temperature, T=0. 
However, it is observed that at finite temperature both isovector
and isoscalar pairing correlations are enhanced with increasing deformation,
which contradict the mean-field predictions.
It is also demonstrated that the pair correlations, which are quenched 
at T=0 and high rotational frequency  re-appear at finite temperature. The 
changes in the individual multipole pairing fields as a function of 
rotation and deformation are analyzed in detail.  
\end{abstract}


\maketitle

The interplay between the deformation forces and the pairing correlations
play a fundamental role in our understanding of rotational 
nuclei \cite{bmii,rs80}. The deformation leads to regular band structures 
and can be measured from the observed transition probabilities. Most of the
nuclei with more than a few valance neutrons and protons 
are known to be deformed. 
The importance of the pairing correlations is inferred, for instance, from the 
odd-even mass differences and the moments of inertia of the rotational states.
These correlations are known to decrease with increasing rotational frequency
along the yrast line.
The deformation and the pairing forces are known to have opposite tendencies.

The dependence of the pairing correlations on temperature is mainly derived
from the mean-field grand-canonical ensemble theory. The experimental
data at finite temperature (excitation energy) is difficult to analyze because
of the quasi-continuum $\gamma-$ray spectrum \cite{doss96}. The mean-field 
models demonstrate
that the pairing correlations drop with increasing temperature and depict
a phase transition from the paired to the un-paired configuration, where the
pairing has completely vanished. The 
mean-field based approach is appropriate
for a macroscopic system, for example, a metallic superconductor. The
experimental data for a bulk superconductor quite clearly depict a
sharp transition as a function of temperature as predicted by the mean-field
models. The application of the mean-field theory to finite mesoscopic systems
also predict a sharp phase transition \cite{good81,good84,er93} 
as in the case of macroscopic systems.
However, the experimental data and the exact solutions of some toy models
don't depict any sharp transition and show a smooth phase transition
from the paired to the un-paired phase \cite{drr93,bfv03,aer01}. The 
application of the mean-field
theory to finite mesoscopic systems is quite inadequate as
it does not contain the fluctuations. The fluctuations, as is known
from statistical mechanics \cite{landau_stat}, are inversely 
proportional to the square-root
of the particle number and become important for a finite system.
In particular, at finite temperature, the mean-field Slater determinant
is a mixture of both even and odd particle numbers and the fluctuations 
become exceedingly important. At zero temperature, the mean-field wavefunction
is mixture of only even or odd particle numbers. 

The projected mean-field theory, which incorporates fluctuations, is
now available at zero temperature \cite{fo96,sr2000}. Recently, there 
has been an attempt to
formulate the projection theory at finite temperature. A partial particle
number projection at finite temperature in which the particle number parity 
projection (even or odd particle number is projected out) has been 
recently studied in ref. \cite{bfv99}. It has been shown, for
a system with odd particle number, that
the pairing correlations which are quenched at zero temperature
and finite magnetic field (rotational frequency) re-appear at finite 
temperature. This interesting and quite unexpected observation was further
 investigated in an exactly solvable model \cite{bfv03}. It was found that 
the pair correlations which were 
quenched at zero temperature  either by an external magnetic field in a small 
superconductor or by rapid rotation in a nucleus, re-appear at finite 
temperature. These ``temperature induced pair correlations'' were noted both 
for even and odd particle numbers. 
In refs. \cite{bfv99,bfv03} the temperature dependence of 
only the monopole  pairing field
has been investigated. 
It is quite interesting to study the temperature dependence
of the higher multipole pair fields for nuclei, where these 
higher fields are known to be important. In particular, it is
quite instructive to investigate 
the temperature and rotational dependence of the isoscalar pairing
field, which has attracted considerable interest in 
recent years \cite {sw2000,dean}.
  
The purpose of the present work is to study the rotational and 
deformation dependence of isovector and isoscalar pairing correlations 
at finite temperature in an exactly solvable model. It is demonstrated that 
the isovector pairing correlations decrease with increasing
deformation at T=0 and the isoscalar pairing field remains constant with
increasing deformation. However, at finite temperature the pairing
correlations depict an increase with increasing deformation, which is totally
unexpected and contradict the mean-field predictions for the specific
model investigated in this paper. 

The exactly solvable model Hamiltonian employed in the present study 
consists of a deformed
one-body term ($h_{def}$) and a scalar two-body delta-interaction ($V_{2}$) 
\cite{s198}. The one-body term is the familiar Nilsson mean-field
potential which takes into account of the long-range part of the
nucleon-nucleon interaction. The residual short-range interaction is
specified by the delta-interaction. The model Hamiltonian is given
by 
\begin{equation}
\hat H=\hat h_{def}+\hat V_{2},  \label{E01}
\end{equation}
where, 
\begin{equation}
\hat h_{def}=-4\kappa \sqrt{\frac{4\pi }{5}}\sum_{i,j}<j|Y_{20}|i>
\delta_{\tau_i \tau_j} \delta_{m_i m_j} c_j^{\dagger}c_i,  \label{E02}
\end{equation}
and 
\begin{eqnarray}
\hat V_{2}& = &{\frac{1}{4}} \sum_{ijkl} <ij|v_a|kl> 
                 c_i^{\dagger}c_j^{\dagger}c_l c_k \nonumber \\ 
     & = &{\frac{1}{2}}\sum_{JMtt_z}v_{Jt}^{{}}A_{JM;tt_z}^{\dagger }
                          A_{JM;tt_z}^{{}},
\label{E03}
\end{eqnarray}
with $A_{JM;tt_z}^{\dagger }=
(c_{j1/2}^{\dagger }c_{j1/2}^{\dagger })_{JM;tt_z}$ and $%
A_{JM;tt_z}^{{}}=(A_{JM;tt_z}^{\dagger })^{\dagger }$. The labels $i,j,....$
in the above equations denote the magnetic quantum number and the isospin
projection quantum number $\tau$ ($\tau=1/2$ for neutrons and $\tau=-1/2$
for protons). For the antisymmetric-normalized two-body 
matrix-element ( $v_{Jt}$ ), we use
the delta-interaction, which for a single j-shell is given by \cite{gb} 

\begin{eqnarray}
v_{Jt} &=& -G{\frac{(2j+1)^{2}}{2(2J+1)}}\biggr\{ \left[ 
\begin{array}{ccc}
j & j & J \\ 
\frac{1}{2} & -\frac{1}{2} & 0
\end{array}
\right] ^{2} \nonumber \\
&&  + {\frac {1} {2}} [1+(-1)^{t}] \left[ 
\begin{array}{ccc}
j & j & J \\ 
\frac{1}{2} & -\frac{1}{2} & 0
\end{array}
\right] ^{2} \biggr\}
\label{E04}
\end{eqnarray}
where the symbol $[~~]$ denotes the Clebsch-Gordon coefficient.
The deformation energy $\kappa $ in Eq. (\ref{E02}) is related to the
deformation parameter $\beta $ \cite{s198}. For the case of $f_{7/2}$
shell, $\kappa $=2.4 approximately corresponds to $\beta =0.25$. 

In the present work, the pairing correlations have been calculated
using canonical ensemble since the exact solutions have well defined
particle number. The average value of a physical quantity ``F'' in canonical
ensemble is given by \cite{landau_stat}
\begin{equation}
<<F>> = \sum_{i} F_i e^{-E_i/T}/Z,
\end{equation}
where,
\begin{eqnarray}
Z     &=& \sum_{i} e^{-E_i/T} \nonumber \\
\hat H |i> &=& E_i |i> \nonumber \\
F_i   &=& <i|\hat F|i>
\end{eqnarray}
The pairing correlations for the $(Jt)$ multipole field is calculated
in the canonical ensemble as
\begin{eqnarray}
E_{Jt}(\rm pair) &=& v_{Jt} {\sqrt {(2J+1)(2t+1)}} \\ \nonumber
&& \biggr( << (A_{Jt}^{\dagger } \times \tilde {A}_{Jt})_{00;00}>> \\ \nonumber
& &-~ {}_{0}<< (A_{Jt}^{\dagger } \times \tilde {A}_{Jt})_{00;00}>>_0 \biggr),
\end{eqnarray}
where the uncorrelated contribution denoted by ${}_{0} <<>>_0$ 
has been subtracted \cite{bfv03}.
The pair-gap for the multipole field $(Jt)$ is then calculated 
through the expression
\begin{equation}
E_{Jt}(\rm pair) = \frac {\Delta_{Jt} \Delta_{Jt}^{\ast}} { v_{Jt}}.
\end{equation} 

The model Hamiltonian has been solved exactly for protons and neutrons
in $f_{7/2}$ subshell. We have considered this subshell as in our
earlier studies \cite{s198}, since the dimensions of the matrices to
be diagonalized are tractable. The results of the total t=1 and 0 pair-gaps  
as a function of rotational frequency for 4-protons and 4-neutrons 
are presented in Fig. 1. For temperature, T=0, both $\Delta_{t=1}$
and $\Delta_{t=0}$ depict changes as a function of rotational frequency.
The decrease in the pair-gaps occur in steps for $\kappa=0$ and for
$\kappa = 3$ the drop occurs in a smoother manner. This decrease in the
pairing correlations
is due to crossing of the aligned configurations with the ground-state
band. The yrast band at low frequencies is a paired state and this
band is crossed by the aligned bands which have reduced pairing correlations.
These aligned bands become favored with increasing rotational frequency.
For higher temperatures, it is noted that the pairing drops smoothly.
The reason for the smooth decrease is that at higher temperatures 
there are crossings of many bands that occur at slightly different
frequencies. The average of these crossings then gives rise to a smoother
drop \cite{bfv03}.

The drop in the t=1 channel in Fig. 1 is similar to
what was found in ref. \cite{bfv03} for the monopole pairing among
identical particles. As discussed below, the decrease of the total
t=1 gap reflects decrease of the monopole term, which is the dominating
part.  The pair correlations in t=0 channel, on the other hand, are 
only moderately reduced. This has been discussed before 
\cite{sw2000,good99,fs2000} and is 
expected as the t=0 pairing should be favored by rotation
(the isoscalar pairs carry angular momenta). However, we shall demonstrate
below that the dipole field (J=1) of t=0 pairing channel is reduced by the 
rotational alignment as the monopole field (J=0) of t=1 pairing, which results
in the moderate decrease of the total t=0 pair correlation.
For the J=1 pair field, the angular momenta of the nucleons are 
nearly anti-parallel and, therefore, rotational alignment of a pair of 
nucleons breaks both the dipole J=1 and the monopole J=0 pairs. 

It is also evident from Fig. 1 that with increasing temperature, the
pairing correlations at higher frequencies become stronger. 
This is quite
unexpected since mean-field theory predicts vanishing of pair
correlations both with increasing rotational frequency and temperature.
In order to demonstrate it more clearly, the pair-gaps are plotted
in Fig. 2 as a function of temperature for three different values
of the rotational frequency.  The upper
panel of Fig. 2 gives the results of pairing correlations for the 
spherical case with $\kappa=0$. For $\hbar \omega = 0$, the pair 
correlations drop with increasing temperature, but it is to be 
noticed that this drop
is very smooth and they do not tend to zero with increasing
temperature. In the mean-field BCS theory, the pair correlations 
depict a sudden transition from finite $\Delta$ to zero $\Delta$.
The non-vanishing of the total pair-gap will be shown later due to non-zero
values of higher-multipole pairing fields. For the $\hbar \omega = 2$ 
and 4, it is seen from Fig. 2 that the pair-gaps increase with
increasing temperature. 
The pairing correlations, which are quenched
for these cases at low temperatures, re-appear at finite temperature. This
result totally contradicts the well known notion based on 
the mean-field models that pair correlations are always quenched with
increasing temperature.  It has first been reported   
in ref. \cite{bfv99} that the number-parity
projection leads to re-appearance of the pairing correlations for
an odd-particle system at finite temperature. In the case of
even-particle system, however, no such re-appearance was found.
This result appears difficult to comprehend as it is expected that the pair
correlations should be stronger in an even system as compared to
an odd-system and, therefore, the re-appearance should be more pronounced
in an even-system. The reason for this inconsistency is 
due to the approximate partial particle-number projection 
( number-parity) performed in ref.\cite{bfv99}. 
Studying the exact solutions  of a  system of fermions 
in a single j-shell interacting via a monopole pair force, Ref. \cite{bfv03}
 found that the
pair correlations that are quenched at zero temperature by 
rotation or a magnetic field re-appear at finite temperature {\it both}
for even and odd particle numbers.

The reason for the above re-appearance
of the pairing correlations at finite temperature can be understood
by first noting that only the occupations of the time-reversed states
close to the Fermi level contribute to the pairing correlations. At
higher rotational frequencies, the particles near the Fermi surface
occupy aligning states, which have large angular-momenta along 
the rotational axis. 
The occupation of these aligning states close to the Fermi surface block
the pairing correlations. However, with increasing temperature these
aligning particles are promoted to higher excited states and the states
close to the Fermi surface can now be occupied in time-reversed form
and consequently increasing the pairing correlations with increasing
temperature. 

In order to critically investigate the behavior of the pair-gaps
as a function of rotational frequency and temperature, the results
of the individual multipole pair-fields are presented in Figs. 3 and 4. 
In Fig. 3, the results are shown for the individual pairing
fields as a function of the rotational frequency at zero temperature, T=0. 
It is evident from this figure that for $\kappa=0$, the changes 
in the total isovector pair-gaps (t=1) noted in Fig. 1 are primarily 
due to the changes in the monopole pairing-field 
($\Delta_{t=1}(J=0))$. The other contributing pair multipoles of J=2, 4 and 6
are quite small and don't depict any changes as a function of rotational 
frequency. For the higher multipoles the two nucleons are not in 
nearly or in fully anti-parallel states and, therefore, these pairs are 
comparatively less affected by rotational alignment.  
For the case of isoscalar pairing-gap, it is seen that the dipole
field drops with rotational frequency. This drop is quite similar to that
seen on the left hand side of the figure for the monopole case. However, the
magnitude of the changes for the dipole field are lower than
the monopole field. It is also noted from the figure that the J=7
multipole field of the isoscalar pair-gap increases with rotational
frequency which is expected, because the angular momenta of the two nucleons
are parallel.  The net result is that the total $\Delta_{t=0}$ shows lesser
variation with rotational frequency as compared to the t=1 pair field.
This strong reduction of the t=1 correlations and the weak change of
the t=0 correlations has also been found in a realistic Cranked Shell Model
Monte Carlo (CSMMC) study for $^{74}$Rb \cite{dean}. In this study, the pair
correlations were studied at zero temperature
and it would be interesting to perform CSMMC calculation at finite temperature
to confirm the re-appearance of the pairing correlations as obtained
in the present model study.

For the deformed case, shown in the lower panel of Fig. 3, there is a 
smoother drop with increasing rotational frequency. The reason for 
this smoother drop is that the wavefunctions in the deformed case
do not have well defined angular-momentum, which leads to a 
smoothing out of the band crossings. 
In the spherical case, the band crossings are sharp, since the wavefunction
has well defined angular momentum. As a consequence, the pairing field 
depicts sudden changes with rotational frequency. It is also noted from Fig. 3
that the t=1 pair field is smaller in the deformed case as compared to
the spherical case. However, the t=0 pairing field is quite similar in the
two cases. 

The results for $\hbar \omega
=4$ are shown in Fig. 4. It is now clearly evident from this figure, that
the re-appearance of the pairing correlations apparent in Fig. 2 are due to the
re-emergence of the monopole (dipole) pair-correlations in 
the t=1(0) channel. The monopole field is zero at low temperatures and becomes
non-zero at about T=0.6 MeV and then shows a smooth increase with temperature.
The dipole field on the right hand panel of Fig. 4 is almost constant at
lower temperatures, but starts increasing at around T=3 MeV. For finite
deformation, the monopole pair-field is constant till T=3 MeV and then
shows an increase.

In our single j-shell model, the single particle levels spread out with 
increasing deformation. For a given interaction strength, 
the pair correlations are expected to decrease with increasing distance 
between the time-reversed states among which the pairs can scatter. 
In order to investigate it, the pair correlations are presented 
in Figs. 5 and 6 as a function of the
deformation parameter, $\kappa$. The  
pair-gaps in Fig. 5, show the expected deformation dependence at 
temperature  T=0. The isovector
correlations drop with increasing deformation and the isoscalar
pair correlations remain constant. However, for finite
temperature, both the isovector and the
isoscalar pair correlations slightly increase with deformation. For example,
the isovector, $\Delta_{t=1}$ increases from 4.4 MeV at $\kappa=0$
to a value of 5.0 MeV at $\kappa=8$MeV. This increase in the pair
correlations with deformation is quite unexpected and contradicts
the mean-field analysis. The mean-field studies predict vanishing of pair 
correlations both with temperature \cite{good84} and 
deformation \cite{srlr02} for the single j-shell model.

The dependence of individual pair fields as a function of deformation
is presented in Fig. 6. It is 
evident from this figure that the increase and decrease 
in the isovector pair field is
essentially determined by the monopole component. The monopole pair
field decreases with deformation at temperature, T=0 and shows an
increasing trend for T=3 and 6 MeV. The other multipole components
J=2, 4 and 6 are quite small and don't depict any significant changes
with deformation. For the isoscalar pair field, it is noted that 
all the components are constant with deformation at temperature, T=0.
At higher temperatures of T=3 and 6, the isoscalar pair fields of 
J=1 and 7 slightly increase with deformation. 

It is to be noted that the real
deformation dependence of the pair correlations 
may be different from our simple model, for which the the deformation leads
only to an increase of the distance between the single particle levels. 
In a realistic potential, the level density is an oscillating function of 
the deformation and the strength of the interaction matrix elements
also depends on the deformation. Nevertheless, our results  seem to suggest
that for low level density systems, the pair correlations may be
slightly enhanced at finite temperature as compared to zero temperature.

In conclusion, we have investigated in an exactly solvable model
the rotational and the deformation dependence of the isovector 
and isoscalar pair correlations at finite temperature. The results
at higher temperatures have been shown to be quite surprising. It has
been noted that the pair correlations re-appear at finite temperature
after they have been quenched at zero temperature and high rotational
frequency. It has shown that the monopole and the dipole pair fields are
responsible for this re-appearance. It has been also observed that the 
pair correlations increase with deformation at finite temperature, which
contradicts the mean-field predictions.

\newpage

\begin{figure*}
\vspace{2cm}
\begin{center}
\includegraphics{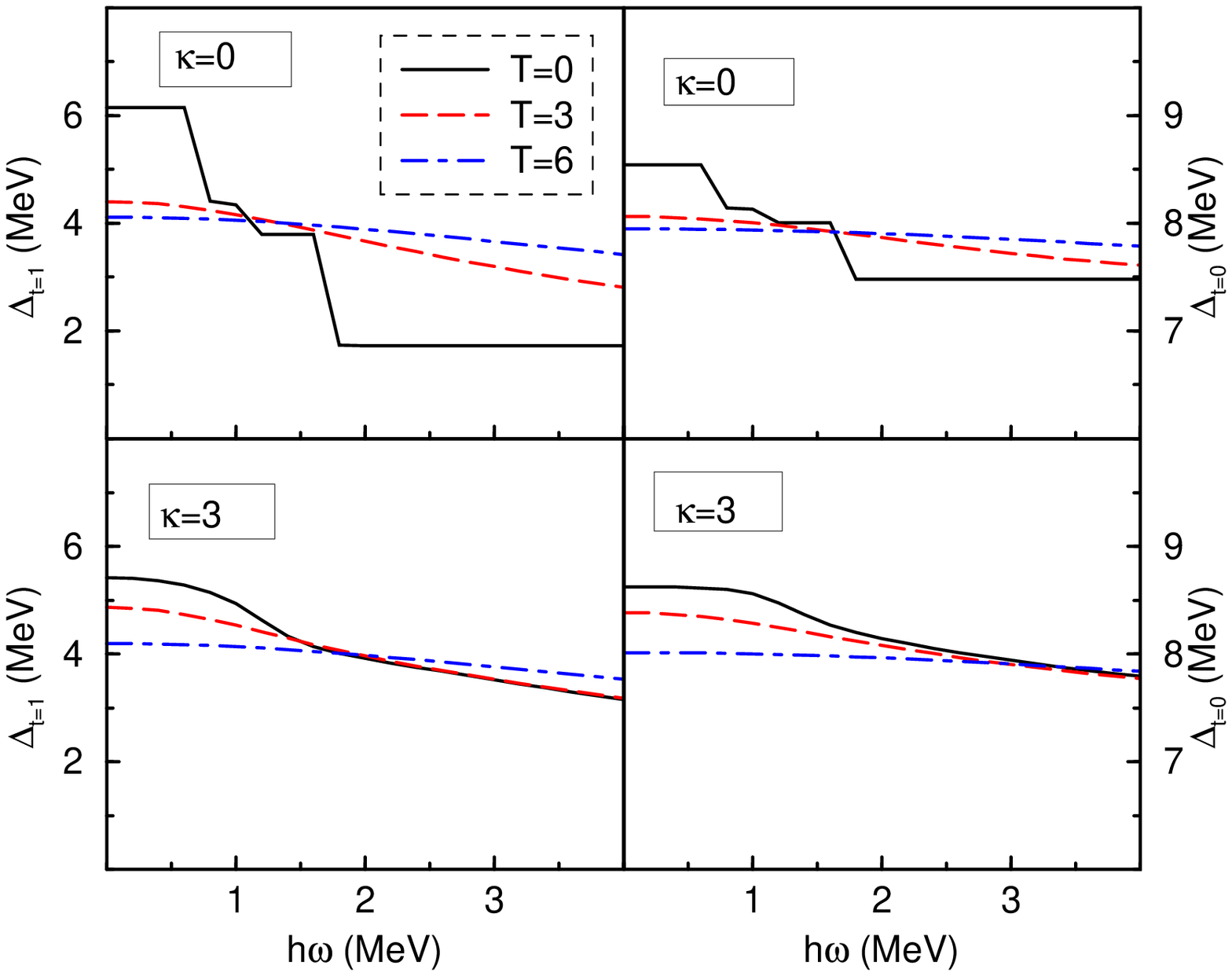}
\caption{
Results of the total isovector ($\Delta_{t=1}$) and isoscalar
($\Delta_{t=0}$) pair-gaps are plotted as a function of rotational frequency
for three different temperatures of T=0, 3 and 6 MeV. The results for spherical
($\kappa = 0$ MeV) and deformed ($\kappa = 3$ MeV) nuclei are shown in upper 
and lower panels, respectively.}
\label{figure.1}
\end{center}
\end{figure*}

\begin{figure*}
\vspace{2cm}
\begin{center}
\includegraphics{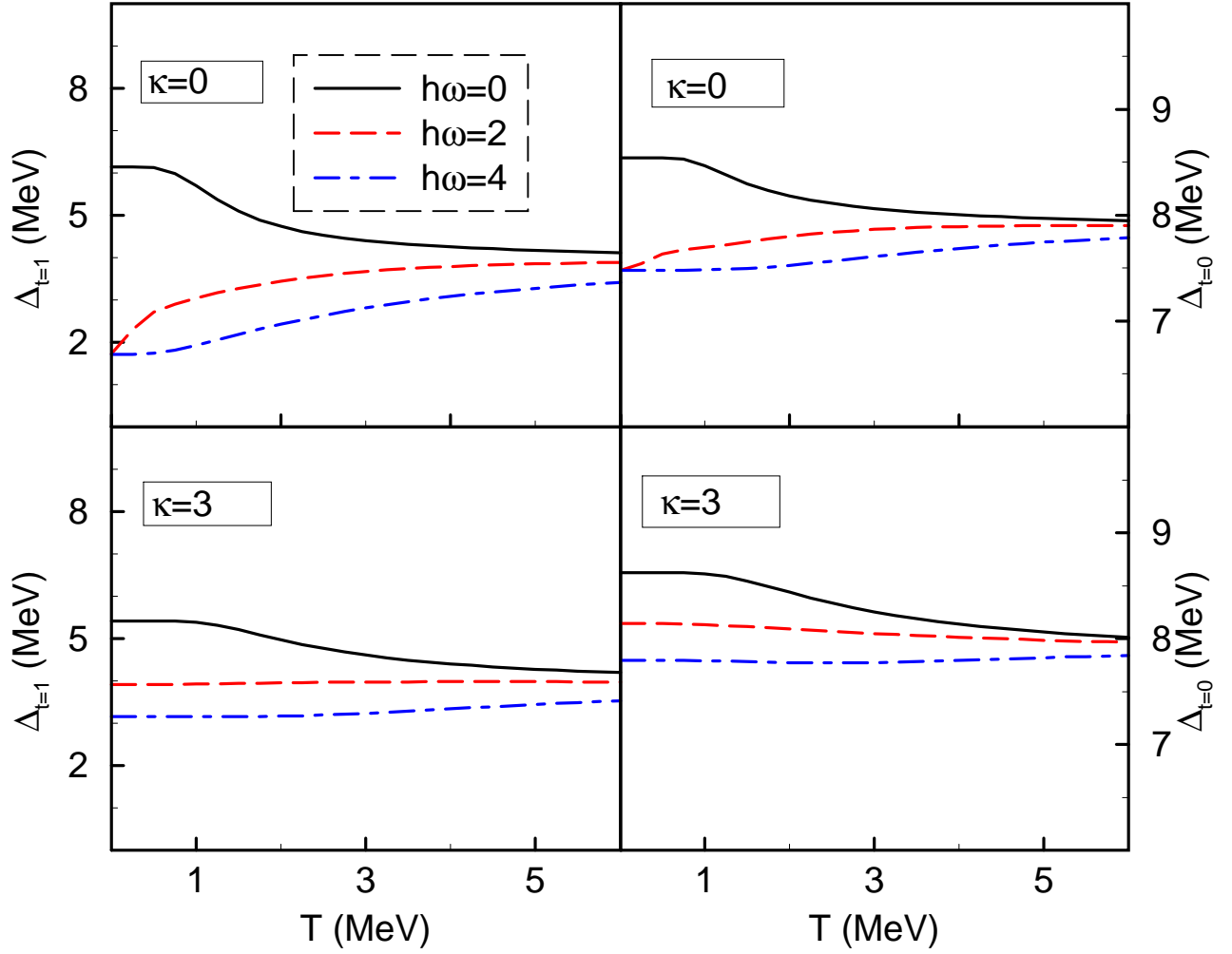}
\caption{
Results of the total isovector ($\Delta_{t=1}$) and isoscalar
($\Delta_{t=0}$) pair-gaps are plotted as a function of temperature
for three different rotational frequencies of $\hbar \omega =$0,
2 and 4 MeV. The upper panel shown the results for $\kappa=0$ and
the lower panel depicts the results for $\kappa = 3$ MeV. 
}
\label{figure.2}
\end{center}
\end{figure*}

\begin{figure*}
\vspace{2cm}
\begin{center}
\includegraphics{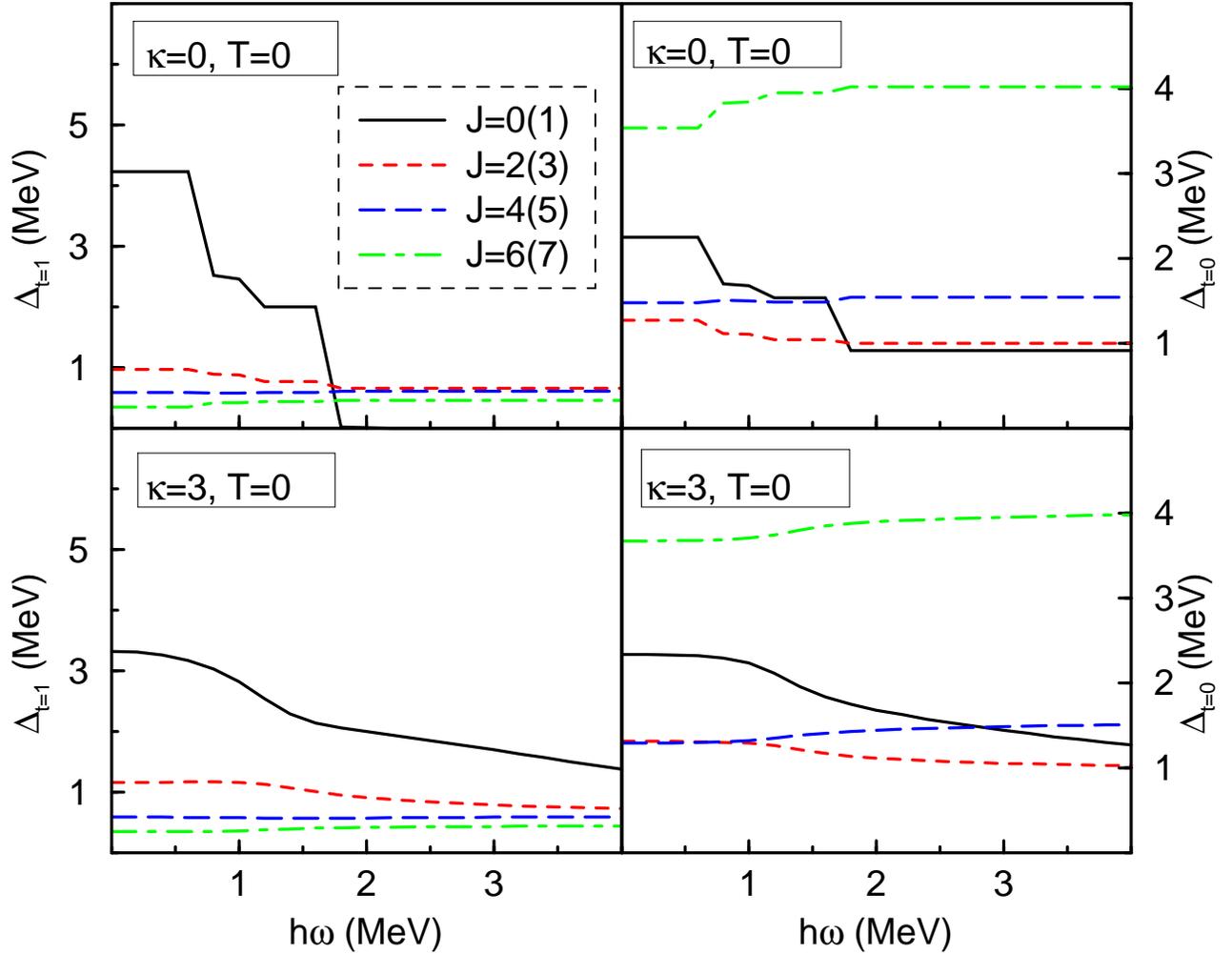}
\caption{
Results of the individual contributing pair fields of 
J=0(1), 2(3), 4(5) and 6(7) for the isovector (isoscalar) are
plotted as a function of the rotational frequency for temperature,
T=0.
}
\label{figure.3}
\end{center}
\end{figure*}

\begin{figure*}
\vspace{2cm}
\begin{center}
\includegraphics{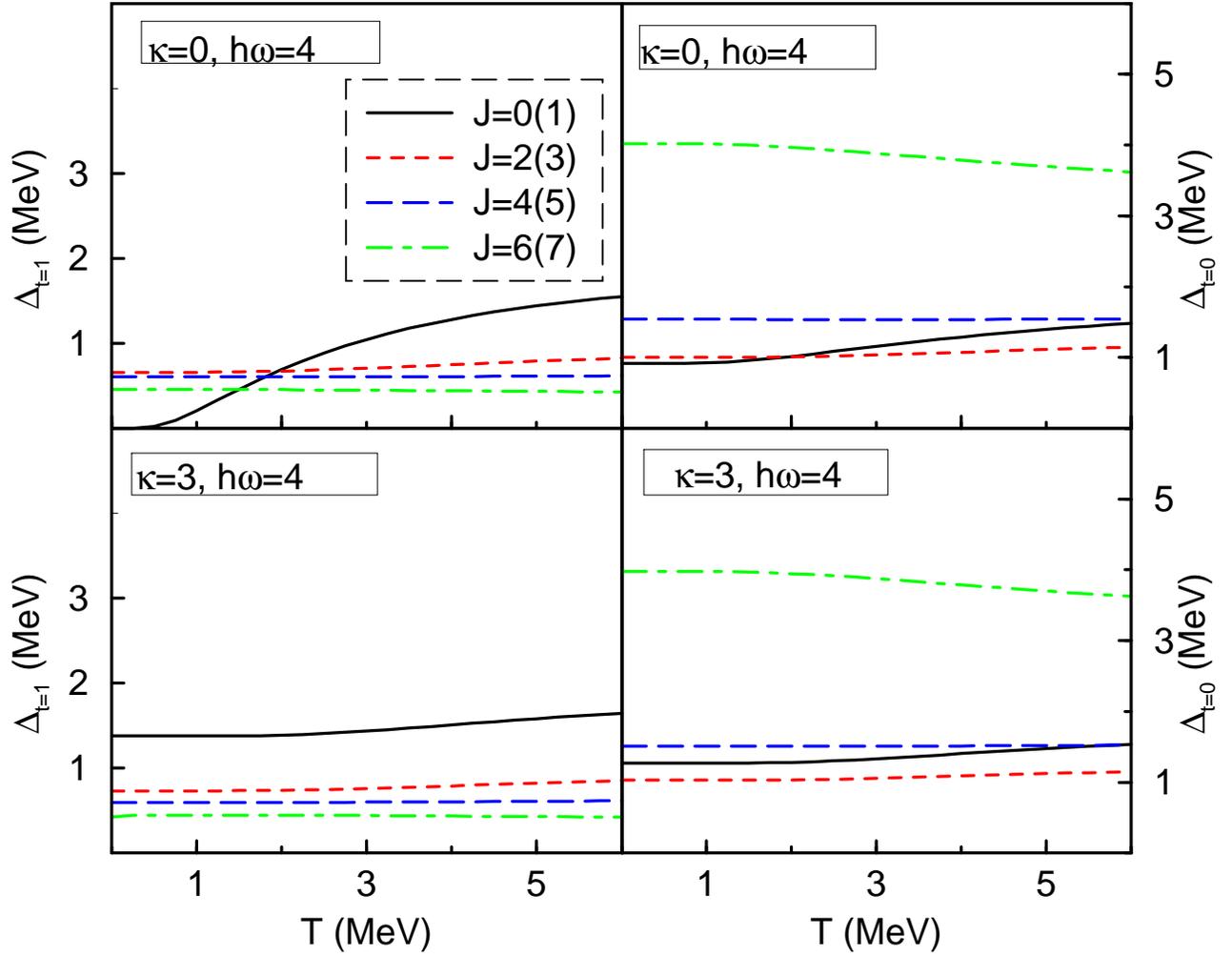}
\caption{
Results of the individual contributing pair fields of 
J=0(1), 2(3), 4(5) and 6(7) for the isovector (isoscalar) are
plotted as a function of temperature for the rotational frequency
of $\hbar \omega =$ 4 MeV.
}
\label{figure.4}
\end{center}
\end{figure*}

\begin{figure*}
\vspace{8cm}
%\begin{center}
\centering
\includegraphics{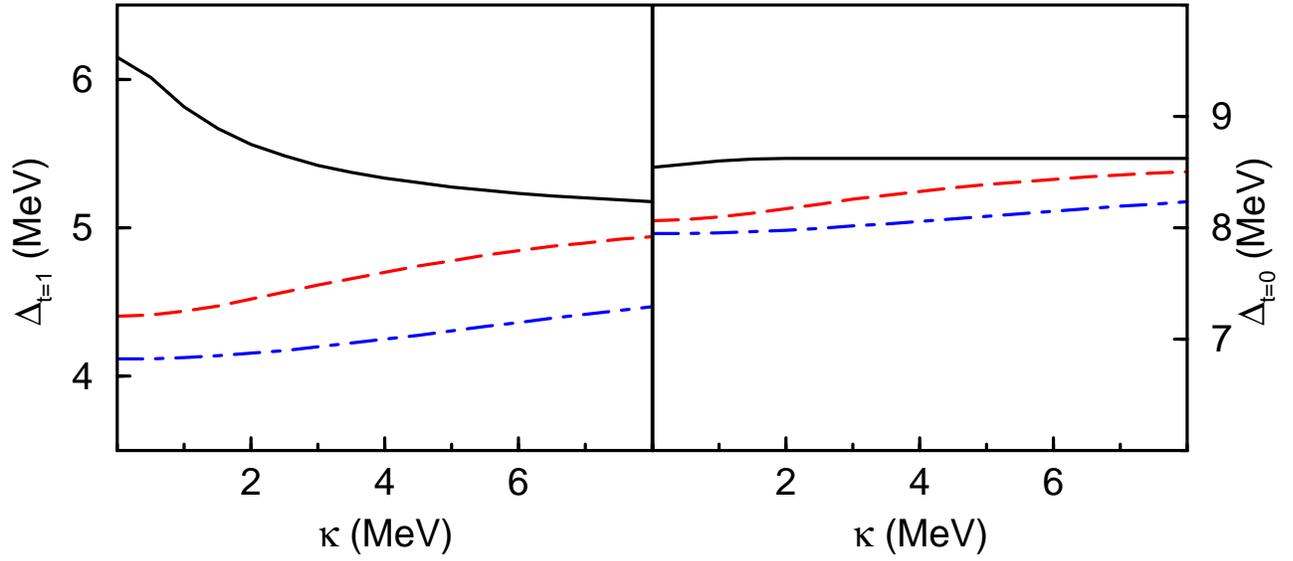}
\caption{
Deformation dependence of the total isovector and isoscalar
pair correlations for three temperatures of T= 0, 3 and 6 MeV, shown
by solid, long dashed and dot-dashed lines, respectively.}
\label{figure.5}
%\end{center}
\end{figure*}

\begin{figure*}
\vspace{3cm}
\centering
%\begin{center}
\includegraphics{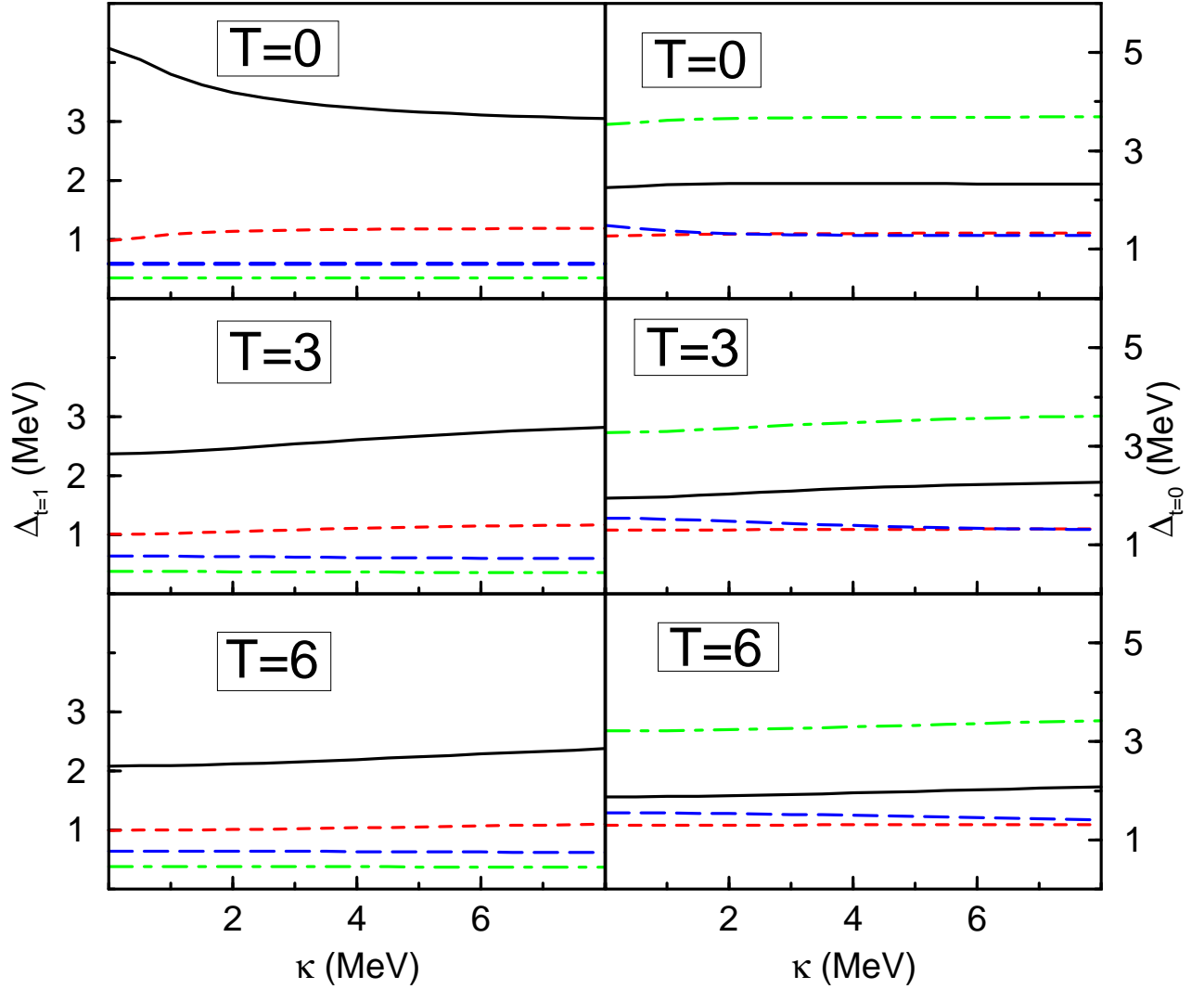}
\caption{
Deformation dependence of the individual contributing 
pair fields of J=0(1), 2(3), 4(5) and 6(7) for the
total isovector (isoscalar) correlations at temperatures 
of T= 0, 3 and 6 MeV. J=0(1) is shown by solid line, J=2(3) by dashed line, 
J=4(5) by long dashed line and J=6(7) by dot-dashed line.}
\label{figure.6}
%\end{center}
\end{figure*}

\end{document}